\documentclass[11pt,a4paper]{article}
\usepackage{jheppub}
\usepackage{gensymb}
\usepackage{ae} 
\usepackage{aecompl}
\usepackage{bm} 
\usepackage{slashed}
\usepackage[usenames,dvipsnames]{color}

\usepackage{amsmath}
\usepackage{amssymb}
\usepackage{epsfig}

\usepackage[utf8]{inputenc}

\usepackage{fontenc}
\usepackage{graphicx}
\usepackage{float}
\usepackage[lofdepth,lotdepth,caption=false]{subfig}
\usepackage{color}
\usepackage{booktabs}
\usepackage{tabularx}
\usepackage{multirow}
\usepackage{longtable}
\usepackage{dcolumn}
\usepackage{rotating}%
\usepackage{booktabs}
\usepackage{tabularx}
\usepackage{makecell} 

\def\nue{{\nu_e}}

\def\numu{{\nu_{\mu}}}

\newcommand{\eg}{{\it e.g.}}
\newcommand{\ie}{{\it i.e.}}
\newcommand{\etc}{{\it etc.}}

\newcommand{\beq}{\begin{equation}}
\newcommand{\eeq}{\end{equation}}
\newcommand{\beqa}{\begin{eqnarray}}
\newcommand{\eeqa}{\end{eqnarray}}

\newcommand{\ta}{\theta_{12}}
\newcommand{\tb}{\theta_{13}}
\newcommand{\tc}{\theta_{23}}
\newcommand{\da}{\delta_{\text{CP}}}
\newcommand{\ldm}{\Delta m_{31}^2}
\newcommand{\sdm}{\Delta m_{21}^2}

\newcommand{\emax}[1]{E^{\text{max}}_{#1}}

\newcommand{\mue}{\nu_\mu \rightarrow \nu_e}
\newcommand{\muebar}{\bar{\nu}_\mu \rightarrow \bar{\nu}_e}
\newcommand{\mumu}{\nu_\mu \rightarrow \nu_\mu}

\newcommand{\pmue}{\text{P}_{\mu e}}
\newcommand{\pmumu}{\text{P}_{\mu\mu}}
\newcommand{\pmuebar}{\bar{\text{P}}_{\mu e}}

\newcommand{\acp}{\Delta P_{\mu e}}

\newcommand{\chisq}{\Delta\chi^2}

\hypersetup{
  colorlinks,
  citecolor=red,
  linkcolor=blue,
  urlcolor=blue}

\title{
Neutrino Oscillation Prospects with a Dual-Baseline Beam from BNL to SNOLAB and SURF}

  \author[a,b,1]{Nishat Fiza,}
 \author[a,b,2]{Mehedi Masud,}    
 \author[b,c,3]{Kim Siyeon,}
 \author[d,4]{Guang Yang}

\affiliation[a]{Sección Física, Departamento de Ciencias, Pontificia Universidad Católica del Perú, Av. Universitaria 1801, Lima 15088, Perú} 
\affiliation[b]{High Energy Physics Center, Chung-Ang University, Seoul 06974, Korea}  
\affiliation[c]{Department of Physics, Chung-Ang University, Seoul 06974, Korea}
\affiliation[d]{Brookhaven National Laboratory, Upton, NY 11973, USA}

\emailAdd{nishatfiza2024@gmail.com} 
\emailAdd{mmasud@pucp.edu.pe}
\emailAdd{siyeon@cau.ac.kr}  
\emailAdd{gyang1@bnl.gov}      
\abstract{
The Electron–Ion Collider (EIC) is a next-generation accelerator primarily designed to study the internal structure of nucleons through high-precision electron–hadron collisions. In this work, we explore the feasibility of employing a 1 MW fraction of the EIC proton beam to generate a high-intensity GeV-scale neutrino beam for long-baseline oscillation studies. We have simulated proton–target interactions and optimize the resulting neutrino fluxes for water-based liquid scintillator (WbLS) detectors located at distinct baselines of 900 km and at 2900 km. 
Oscillation analyses performed with GLoBES show that extended baselines allow access to multiple oscillation maxima, significantly enhancing sensitivity to leptonic CP violation. The study also examines the interplay between matter effects and the intrinsic CP violating phase in shaping observable asymmetries.
\textcolor{black}{We note that simplified systematics and no backgrounds are used in this analysis to establish the baseline physics potential.}
These results suggest that the EIC proton beam could provide a novel and complementary source for precision neutrino physics, extending the scientific reach of the EIC program.
}
\begin{document}
\maketitle

\tableofcontents
\section{Introduction}\label{sec:intro}                                                  
Neutrinos continue to occupy a central place in the search for physics beyond the Standard Model.
The discovery that they oscillate between flavors, implying non-zero mass and lepton mixing, has been one of the most important breakthroughs in particle physics in the past two decades~\cite{Super-Kamiokande:1998kpq, SNO:2001kpb, nobel2015}. 
Standard 3-flavour neutrino oscillation is governed by two mass-squared differences, $\ldm$ and $\sdm$, the three mixing angles, $\ta, \tb, \tc$ and the CP phase, $\da$~\cite{Pontecorvo:1957qd,Pontecorvo:1957cp,Gribov:1968kq,Maki:1962mu}. 
Oscillation experiments have already determined two distinct mass-squared differences ($\sdm, |\ldm|$) and two mixing angles ($\ta, \tb$), and the global analyses of available neutrino data~\cite{Capozzi:2018ubv, 10.5281/zenodo.4726908, 
deSalas:2020pgw, nufit_globalfit, Esteban:2024eli} is consistent with the standard $3\nu$ scenario. 
However, several fundamental puzzles in the neutrino oscillation sector still remain unresolved. 
Chief among these are the size and nature of CP violation in the lepton sector; 
the ordering of the neutrino mass states (normal versus inverted hierarchy); the correct octant of the atmospheric mixing angle $\tc$, as well as the possibilities of the presence of several non-standard oscillation effects~\cite{SajjadAthar:2021prg, Arguelles:2022tki}. 
Addressing these questions is essential not only for completing our picture of neutrino physics, but also for potentially uncovering deeper mysteries such as the origin of matter-antimatter asymmetry in the universe or the process of generation of (tiny) neutrino mass \etc

The long baseline (LBL) neutrino experiments are poised to shed lights on these unresolved issues. 
Although the data~\cite{MINOS:2020llm, NOvA:2021nfi, T2K:2023smv, T2K:2023mcm, NOvA:2024lti} from the current LBL experiments such as Tokai to Kamioka (T2K)~\cite{T2K:2011ypd} (with a baseline of 295 km), NuMI Off-axis 
$\nu_e$ Appearance (NO${\nu}$A)~\cite{NOvA:2019cyt} (800 km baseline), \textcolor{black}{and the recently concluded Main Injector Neutrino Oscillation Search (MINOS \& MINOS+ with a baseline of 735 km)~\cite{MINOS:1998kez} mildly indicate towards a normal mass ordering and a higher octant for $\tc$, the results are still far from being conclusive. 
Non-accelerator based neutrino experiments also attempt towards resolving these issues. 
Atmospheric data from Icecube~\cite{IceCube:2017lak} and Super Kamiokande~\cite{Super-Kamiokande:2019gzr} hint towards a higher and lower octant for $\tc$}.
The future LBL experiments, such as the Deep Underground Neutrino 
Experiment (DUNE)~\cite{DUNE:2020lwj, DUNE:2020ypp}, 
Tokai to Hyper-Kamiokande (T2HK)~\cite{Hyper-KamiokandeWorkingGroup:2014czz} aim to resolve the outstanding issues by employing more advanced next generation  detection techniques. 
Complementary measurements from alternative baselines and detector technologies are invaluable to fully resolve these ambiguities and to strengthen confidence in discoveries. 
\textcolor{black}{Typically the LBL experiments focus on energies around the first oscillation 
maximum for the $\mue$ oscillation (where $L/E \simeq 500$ km/GeV).  
Additionally, performing measurements at the second oscillation maximum,- one of the goals of the future LBL proposal future LBL project European 
Spallation Source $\nu$ Super Beam (ESS$\nu$SB)~\cite{ESSnuSB:2013dql, ESSnuSB:2023ogw}, where $L/E \simeq 1500$ km/GeV, either using longer baseline or by lowering energy can in principle offer better CP violation sensitivities, which are less affected by systematics~\cite{Huber:2010dx, Budimir_cern_ep2022, Giarnetti:2023pkz, Fanourakis:2025bqi}. 
However, a longer baseline means less statistics, while lowering energy decreases 
the neutrino cross-sections. 
Measurements at second maximum also requires very good energy resolution of the detector~\cite{DeRomeri:2016qwo}. 
Studies exploring the second oscillation maximum at other LBL experiments, including DUNE involve for \eg, mass hierarchy analysis~\cite{Ghosh:2014rna}; octant and CP studies with tunable beams~\cite{Rout:2020emr}; as well as some new physics studies~\cite{Choubey:2020dhw, Chakraborty:2020cfu}.}  

The upcoming Electron-Ion Collider (EIC)~\cite{Accardi:2012qut, Aschenauer:2017jsk, AbdulKhalek:2021gbh, Willeke:2021ymc, AbdulKhalek:2022hcn, Xu:2022pem, Adkins:2022jfp} at Brookhaven National Laboratory offers an opportunity to establish such a complementary program. 
Although the EIC's primary mission is to map the three-dimensional structure of nucleons and nuclei through high-energy electron-proton and electron-ion collisions, its accelerator complex is also capable of delivering high-energy polarized proton beams. 
These beams, when directed at a fixed target, can generate a conventional pion-decay neutrino beam similar to the NuMI facility at Fermilab, but at substantially higher proton energies (up to 275 GeV)~\cite{AbdulKhalek:2021gbh, Willeke:2021ymc, AbdulKhalek:2022hcn}. 
The resulting neutrino flux would be highly collimated and broadband, offering unique advantages for long-baseline oscillation studies.  
The neutrino beam could be directed toward two far detectors, - one at the SNOLAB site in Canada (with a baseline of 900 km), and one at the SURF site in USA (with a baseline of 2900 km)~\footnote{Note that the SNOLAB site is currently set to host the SNO+ experiment~\cite{SNO:2021xpa}, which, in addition to be having a water phase~\cite{SNO:2018ydj} and a liquid scintillator phase~\cite{SNO:2020fhu}, is also proposed to have a hybrid water based liquid scintillator~\cite{SNOLAB:2025vfr}. The SURF site, on the other hand is set to host the above mentioned DUNE project.}.  
The shorter baseline allows \textcolor{black}{measurements of oscillation parameters} near the first oscillation maximum with reduced matter effects. 
The longer baseline allows probing more than one oscillation cycles (which is otherwise difficult in LBL experiments), as well as studying matter induced modifications, - thereby offering a powerful probe to CP violation and mass hierarchy~\footnote{The authors of   \cite{Denton:2024glz} have also studied neutrino oscillation in the context of BNL-SURF baseline, but in the context of a neutrino factory setup at BNL. The readers may also look at \cite{Denton:2025kvy} in order to understand how such a neutrino factory setup can probe non-standard neutrino interactions.}. 
Together, the two baselines allow for simultaneous exploration of different oscillation regimes, capturing both the first and second oscillation maxima with complementary strengths.
Comparisons between detectors at different baselines also enable cross-checks that reduce systematic uncertainties. 
Finally, the EIC’s ability to deliver polarized proton beams could open entirely new avenues, such as studying spin-dependent effects in meson production and their impact on neutrino fluxes which is a possibility not accessible in existing neutrino programs. \textcolor{black}{As a first-order feasibility study, we employ simplified systematics and neglect backgrounds in this analysis.}

In this work, we explore the feasibility and physics potential of such an EIC-based dual-baseline neutrino experiment. 
We begin in Sec.\ \ref{sec:motiv} by outlining the motivation and advantages of using the EIC proton beam for neutrino production. 
Sec.\ \ref{sec:flux_calc} presents flux calculations for the proposed baselines, while Sec.\ \ref{sec:prob} analyses oscillation probabilities and their dependence on matter effects and CP violation. 
Sec.\ \ref{sec:event} discusses expected event spectra at the SNOLAB and SURF detectors. 
Sec.\ \ref{sec:chisq} provides preliminary $\Delta \chi^2$ sensitivity estimates for CP violation. 
We conclude in Sec.\ \ref{sec:summary} with a discussion of the broader implications, design considerations, and future directions for integrating such a neutrino program into the EIC infrastructure.

\section{Motivation}\label{sec:motiv}                                                  
The EIC is a state-of-the-art particle accelerator facility being built to solve long-standing mysteries in nuclear physics. Its primary mission is to precisely map the internal 3D structure of nucleons and nuclei by colliding high-energy polarized electron beams with polarized proton and ion beams. By doing so, it will explore the dynamics of quarks and gluons, the fundamental constituents of visible matter, and seek to understand the origin of the proton's spin and mass.
While designed for electron-ion collisions, the EIC's powerful proton accelerator complex provides capabilities that can be harnessed for a world-class neutrino program. Comparison with the proton beam from the Long-Baseline Neutrino Facility (LBNF) at Fermilab, which powers the DUNE experiment, highlights the unique advantages and complementarities. The key parameters of the two proton beamlines are summarized in Table \ref{tab:beam_comparison_final}~\cite{DUNE:2020lwj, Willeke:2021ymc}.

\begin{table}[h!]
\renewcommand{\theadfont}{\bfseries} 
\centering
\caption{Comprehensive Comparison of LBNF and EIC Beam Parameters.}
\label{tab:beam_comparison_final}
\begin{tabularx}{\textwidth}{l X X X} 
\toprule
& \multicolumn{3}{c}{\thead{Facility and Operating Mode}} \\
 \cmidrule(l){2-4}
\thead{Parameter} & \thead{Fermilab \\ LBNF/DUNE} & \thead{EIC Collider Mode \\ (Stored Beam)} & \thead{EIC Fixed-Target \\ (Hypothetical)} \\
\midrule
Proton Beam Energy & 60--120 GeV & Up to 275 GeV & Up to 275 GeV \\
\addlinespace
Beam Power & 1.2 MW (initial) & N/A (Beam is stored) & $\sim$1 MW (assumed) \\
\addlinespace
POT per Second & $6.25 \times 10^{13}$ & N/A (Protons are stored) & $2.27 \times 10^{13}$ \\
\addlinespace
Beam Structure & Pulsed: 10 $\mu$s pulse per ~1.2 s cycle & $\sim$1160 bunches per s($2.6 \times 10^{11}$ p/bunch) & Pulsed beam needing further design \\
\bottomrule
\end{tabularx}
\end{table}

The EIC offers a significantly higher proton energy (up to 275 GeV) compared to LBNF's maximum of 120 GeV. This higher energy would produce a more energetic and highly collimated neutrino beam, increasing the neutrino flux in the forward direction and extending the physics reach for very long baseline experiments~\footnote{From Table \ref{tab:beam_comparison_final}, we note that the POT/second for the 
proton beam in EIC can be assumed to be $2.6 \times 10^{11} \times 1160 = 3.016 \times 10^{14}$. This number when multiplied by the proton energy of 275 GeV, gives 13.2 MW. In order to use only a part of the main proton beam at EIC for neutrino production we take a conservative assumption of an 1 MW beam.}. While the instantaneous intensity of a single LBNF pulse is much higher, the EIC's near-continuous stream of bunches could potentially deliver a substantial number of protons-on-target (POT) over a given run period. 
\textcolor{black}{The idea of a near-continuous stream here refers to the high-repetition pulsed structure of the proton beam at the EIC, which differs from the lower-duty-factor, widely spaced pulses used at conventional neutrino facilities. 
To convert such a beam into towards a fixed target or decay region for neutrino production, additional beam-handling elements, (for \eg, fast-extraction systems~\cite{4329992, Fraser:2018bcq, Shibata:2018wrf}, RF/microwave-based deflection or bunch-structure manipulation techniques~\cite{987379, Burt:2012hc}) - would need to be implemented. These accelerator-engineering aspects require a dedicated feasibility study and lie beyond the scope of the present paper. Here we focus solely on the baseline physics potential under idealized beam-delivery assumptions.}

Crucially, the EIC will be the world's only accelerator capable of delivering highly polarized proton beams at this energy scale. Repurposing this polarized beam for neutrino production would be an interesting step. It would allow for unique studies of spin-transfer in meson (pion and kaon) production, offering a new tool to probe and constrain the systematic uncertainties related to the neutrino production mechanism. In particular, given the protons in the EIC beamline can be both longitudinal and transverse polarized, there are multiple ways to study the meson production and decay processes through the controlled polarization. For example, the transverse polarized proton beam may result in modulation of the meson angular distribution. The meson angular distribution will directly impact the neutrino energy spectrum given fixed detector location. A transversely movable near detector can further explore the nature of the meson production and decay processes. 

\textcolor{black}{Specifically, the transverse polarization of the proton beam induces a Single Spin Asymmetry (SSA) in the production of secondary mesons ($p^\uparrow C \to \pi X$), - well 
established in, for instance, in the data from Fermilab E704~\cite{FNALE704:1994hmp, FermilabE704:1996mxs},  Relativistic Heavy Ion Collider (RHIC)~\cite{STAR:2003lxu, BRAHMS:2008doi, PHENIX:2014qwb} among others. 
This asymmetry results in a controlled azimuthal modulation of the meson flux entering the decay volume, - giving rise to a left-right angular asymmetry. 
At accelerator energies the pion decay produces neutrinos that are mostly confined within a narrow cone around the pion direction, and thus this asymmetry is expected to survive in the resulting neutrinos. 
By measuring this angular asymmetry variation in the neutrino rate with spin-flip, one can disentangle geometric beamline systematics (which do not flip) from production physics, offering a novel in-situ constraint on the neutrino flux that is unique to the EIC. 
In order to get  an estimate of the constraint, we first note that in our setup below, a $\mathcal{O}(10)$ kt far detector at 2900 km from EIC detects roughly $\mathcal{O}(10^{3})$ neutrino events in a year. 
Hence a suitably placed $1$t near detector at around 1 km from the EIC source will detect roughly $N \sim \mathcal{O}(10^{5}-10^{6})$ events, - offering a statistical uncertainly of approximately $1/\sqrt{N} \sim 0.1\%$. 
Further, we note that the hadron production modeling is one of the principal sources of systematics in the LBL experiments~\cite{Ferrari:2005zk, GEANT4:2002zbu, Yarba:2012ih}.
Thus a model based on SSA predicting an angular asymmetry of, say, $15\%$ (see, for \eg, \cite{E704:1995her, PhysRevD.69.017501, STAR:2008ixi}) can potentially be ruled out by data showing $20.0\% \pm 0.1\%$ with a high statistical significance $\sim 50\sigma$. 
This would, in turn, help constrain the model parameters tightly and bring down the underlying cross-section uncertainties in the hadron production models and the resulting neutrino flux uncertainties.
} This topic worth further studies in detail, which can form a separate good motivation for the proposed project.
 
However, in this work, we will explore the benefit to the neutrino oscillation research from the super-long-baseline with unique $L/E$.
A very specific advantage for the next-generation neutrino oscillation experiments is the broad-band beam. In the past experiments such as T2K and NOvA, they are largely counting experiments comparing the total numbers of $\nu_{e}$ and $\bar{\nu}_{e}$ appearance. With the broad-band beam, the shape information can be exclusively used to gain better sensitivity to the neutrino oscillation parameters, particularly the CP violation phase. 
\textcolor{black}{However, the exact shape of the beam depends on the design of the focusing horn 
(horn length, target position relative to the horn, as well as on horn current, among others) that 
determines how the mesons are transported forward and eventually decay to generate neutrinos. 
The upcoming DUNE with a 1300 baseline from Fermilab to SURF is set to employ the intense 1.2 MW  wide band on-axis neutrino beam covering from near 1 GeV to over 10 GeV neutrino energies with a peak around 2.25-2.5 GeV.}
The baseline from Brookhaven National Lab to SURF is roughly two times than the baseline from Fermilab to SURF, thus the first few $\nu_{e}$ oscillation peaks for the longer baseline are higher in terms of energy. Meanwhile, given the overall higher energy proton at the EIC beamline, we could expect more clear second and even third oscillation peaks for such a super long baseline experiment. In our calculation, we dive into the detail how the EIC baseline and energy could impact the neutrino oscillation parameter sensitivities.    

\section{Flux Calculation}\label{sec:flux_calc}                                                  
We calculate the flux following the neutrino flux generation method in NuMI beam in Fermilab.
A Geant4 model has been carried out to implement the neutrino flux simulation~\cite{GEANT4:2002zbu} following the LBNF beamline design report~\cite{Papadimitriou:2016ksv}.
The 275 GeV protons from the storage ring are dumped into a pipe hitting a target.
The target is assumed to be a long thin graphite with dimension of 1.5 m x 8 mm (length x radius).
The resulting hadrons were focused with six horn current settings ranging from 93 kA to 593 kA with a step of 100 kA. The two horn dimensions are assumed to be 4.2 m x 0.4 m and 3.8 m x 0.48 m (length x radius) respectively.
\textcolor{black}{After the horn focusing of the hadrons, the decay pipe  is assumed to be $\sim$300\,m long and 4\,m in diameter to accommodate meson decays resulting in a neutrino beam having energy upto $\mathcal{O}(10 \text{ GeV})$. 
Muon and hadron absorption is achieved using modular aluminum and steel absorber blocks, and the full system requires appropriate shielding and cooling.
Two baselines are considered in this study, requiring adjustments to the horn orientation and neutrino beam angle to account for baseline length and Earth's curvature. 
For reference, the LBNF beam is directed 5.8$^\circ$ below horizontal to reach the DUNE far detector at 1300\,km. 
Using similar considerations, beam angles of approximately 4$^\circ$ and 13$^\circ$ below horizontal are obtained for the two proposed baselines in our study. 
A 1-ton WbLS near detector is proposed at a distance of 500 m to 1 km downstream of the decay pipe. Detailed optimization and engineering design are beyond the scope of this work.}

\begin{figure}[t]
 \centering
 \includegraphics[scale=0.6]{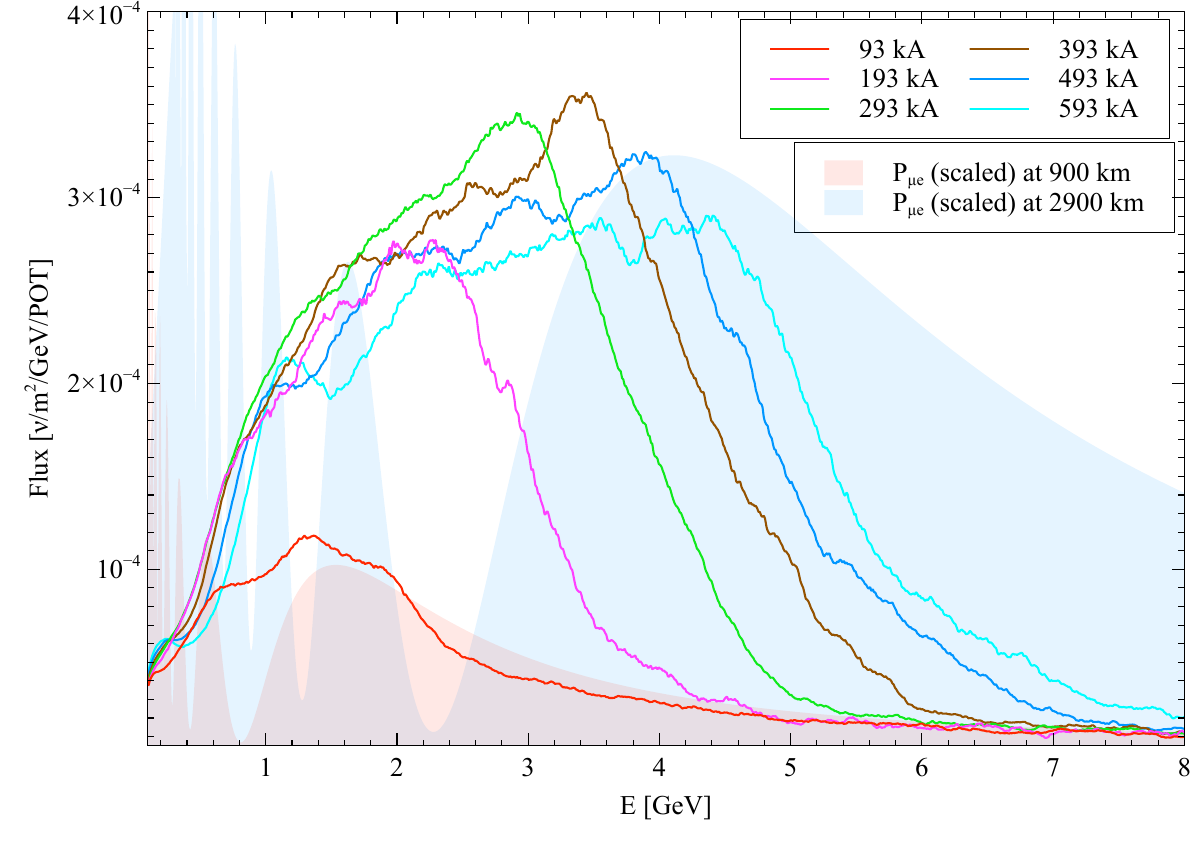}
 \caption{\footnotesize{This shows the $\nu_{\mu}$ flux at 1 km from EIC for six different horn currents (shown in the keys in units of kA). $\pmue$ (with arbitrary scaling) for 900 km and 2900 km are overlaid in red and blue shades respectively.}}
  \label{fig:flux_horncurrent}
 \end{figure}

In Fig.\ \ref{fig:flux_horncurrent} we illustrate the neutrino fluxes (at 1 km from the source) corresponding to 
the six horn current settings.
The red (blue) shaded regions in Fig.\ \ref{fig:flux_horncurrent} show $\pmue$ (estimated using the oscillation parameters in Tab.\ \ref{tab:parameters} and scaled with arbitrary constant for better visibility) for the baselines of 900 km and 2900 km respectively. 
Among the six horn current settings simulated, the ones that resulting energy peaks at the far detector match the first $\nu_{e}$ appearance peaks for the 900 km and 2900 km baselines were selected. 
It turned out the best horn current values are 93 kA and 493 kA for the 900 km and 2900 km baseline respectively.

\begin{figure}[t]
 \centering
 \includegraphics[scale=0.48]{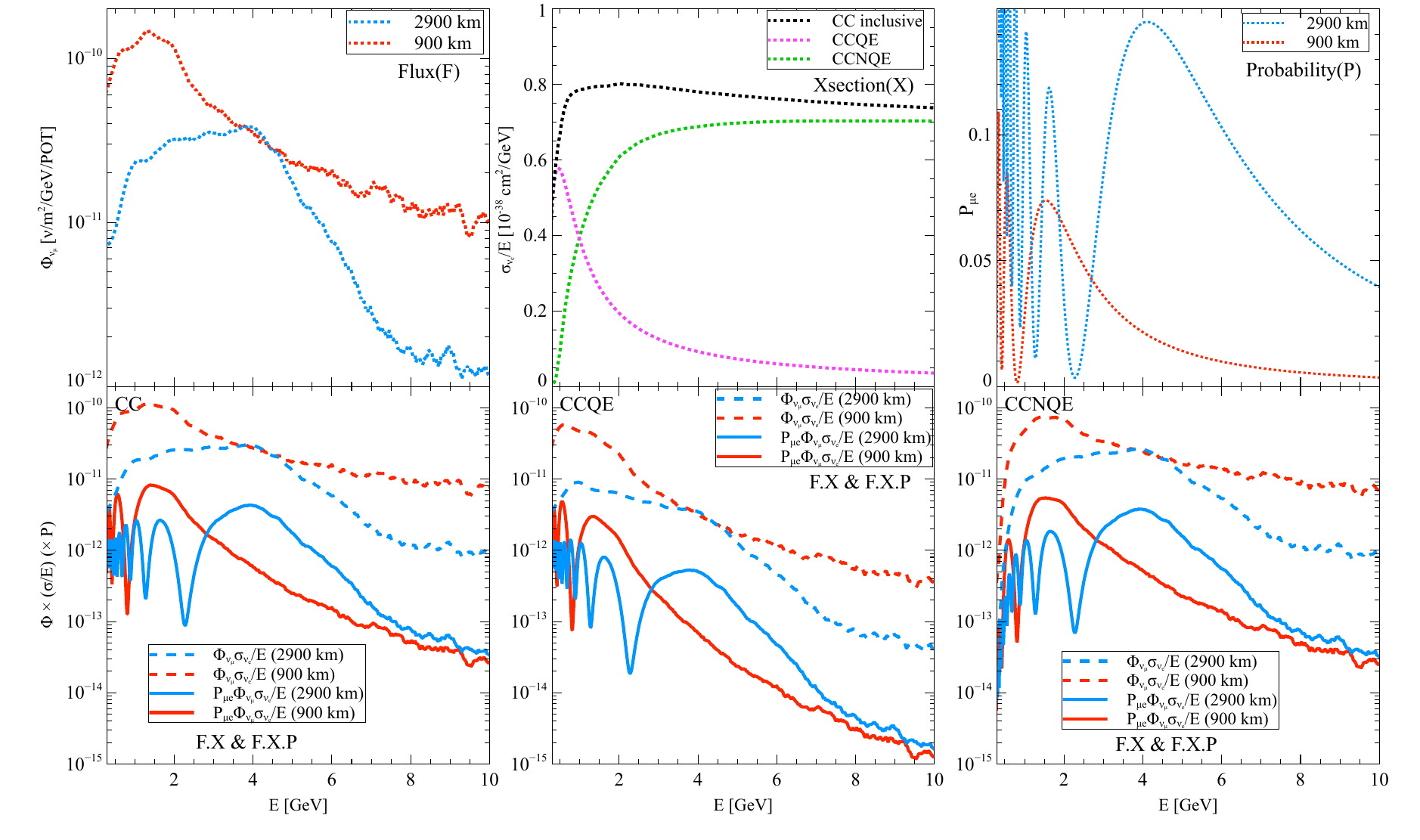}
 \caption{\footnotesize{The top row shows the $\nu_{\mu}$ flux \textcolor{black}{(simply abbreviated as F)} $\Phi_{\numu}$; $\nu_{e}$ cross-section \textcolor{black}{(X)} $\frac{\sigma_{\nue}}{E}$; and the transition probabilities \textcolor{black}{(P)} $\pmue$ at EIC-SURF baseline of 2900 km, and at EIC-SNO baseline of 900 km. The bottom row shows the relevant products of these quantities \textcolor{black}{(\ie, flux $\times$ cross-section or F.X, and flux $\times$ cross-section $\times$ probability or F.X.P)} for three different types of interactions, -namely inclusive charge current (CC), quasielastic charge current (CCQE), non-quasielastic charge current (CCNQE).}}
  \label{fig:flux_xsec}
 \end{figure}

In Fig.~\ref{fig:flux_xsec}, the simulated neutrino fluxes with 93 kA (red) and 493 kA (blue) are shown in the top left panel. The peaks match the first $\nu_{e}$ appearance peak locations.
The top middle panel shows the neutrino interaction cross-section curves for various interaction channels.
The top right panel shows the $\nu_{e}$ appearance oscillation probability for the two baselines.
The bottom panels show the product of the flux and cross section, as well as the product of the flux, cross section and $\nu_{e}$ appearance oscillation probability.
Each bottom panel corresponds to one interaction channel, namely charged-current inclusive, charged-current quasi-elastic and charged-current non-quasi-elastic from the left to right.

The CP violation phase sensitivity with our proposed detector depends on the solid curves. The final sensitivity includes the effect of absolute event rate, energy smearing and systematic uncertainties.
It is obvious that for the 900 km baseline scenario, it is likely we can utilize only one appearance oscillation peak. On the contrary, for the 2900 km baseline scenario, we may use up to three oscillation peaks. The detector smearing may play an important to identify the oscillation peaks.

\section{Probability}
\label{sec:prob}
  \begin{table}[b]
\centering
\scalebox{1.0}{
\begin{tabular}{| c | c | c | c |}
\hline
&&&\\
Parameter & Best-fit-value & 3$\sigma$ interval & $1\sigma$ uncertainty  \\
&&&\\
\hline
$\theta_{12}$ [deg.]             & 34.3                    &  31.4 - 37.4   &  2.9\% \\
$\theta_{13}$ [deg.]    & 8.53  &  8.13  -  8.92   &  1.5\% \\
$\theta_{23}$ [deg.]        & 49.3      &  41.2  - 51.4    &  3.5\% \\
$\sdm$ [$\text{eV}^2$]  & $7.5 \times 10^{-5}$  &  [6.94 - 8.14]$\times 10^{-5}$  &  2.7\% \\
$\ldm$ [$\text{eV}^2$] & $2.55 \times 10^{-3}$   &  [2.47 - 2.63] $\times 10^{-3}$ &  1.2\% \\
$\da$ [rad.]   & $-0.92\pi$   & $[-\pi, -0.01\pi]  \cup [0.71\pi, \pi]$ &  $-$ \\
\hline
\end{tabular}}
\caption{\footnotesize{\label{tab:parameters}
 The values of the oscillation parameters (taken from the global fit analysis in \cite{10.5281/zenodo.4726908, deSalas:2020pgw}) and their uncertainties used in our study.  
If the $3\sigma$ upper and lower limit of a parameter is $x_{u}$ and $x_{l}$ respectively, the $1\sigma$  uncertainty is $(x_{u}-x_{l})/3(x_{u}+x_{l})\%$~\cite{DUNE:2020ypp}.
The neutrino mass hierarchy was assumed to be normal.
}}
\end{table}
In Fig.\ \ref{fig:flux_xsec} (top, right panel) we have already shown $\pmue$ at 900 and 2900 km (calculated numerically using GLoBES by considering the best fit values of the oscillation parameters listed in Table \ref{tab:parameters}). 
In order to understand the features of oscillation at different baselines, we analyse the probability expression in constant density for the $\mue$ channel~\cite{Akhmedov:2004ny}.
\begin{align}\label{eq:prob}
 P_{\mu e} &\simeq \sin^{2}2\theta_{13}\sin^{2}\theta_{23}\frac{\sin^{2}(1-A)\Delta_{31}}{1-A} 
 \nonumber \\
 &+
  \alpha  \sin2\theta_{13} \sin2\theta_{23}  \sin 2\theta_{12} \frac{\sin A\Delta_{31}}{A} 
 \frac{\sin(1-A)\Delta_{31}}{1-A} \cos (\da + \Delta_{31}) \nonumber \\
 &= 
 \frac{\sin(1-A)\Delta_{31}}{1-A} \bigg[\sin^{2}2\theta_{13}\sin^{2}\theta_{23}
 \frac{\sin(1-A)\Delta_{31}}{1-A} \nonumber \\
 &+
  \alpha  \sin2\theta_{13} \sin2\theta_{23}  \sin 2\theta_{12}  
  \frac{\sin A\Delta_{31}}{A} \cos (\da + \Delta_{31})  \bigg],
 \end{align}
 where $\alpha = \frac{\sdm}{\ldm}$ and $\Delta_{31} = \frac{\ldm L}{4E}$. 
 $A = \frac{2\sqrt{2}G_{F}N_{e}E}{\ldm}$ is the matter effect where $N_{e}$ is the electron 
 density along the neutrino propagation length.
\textcolor{black}{For Earth matter, which is mainly composed of medium to heavy nuclei (Si, Fe, Mg \etc), 
 the average value of the ratio of the number density of electrons to that of nucleons is roughly 0.5.} 
 We can then calculate the following quantity as~\cite{Akhmedov:2004ny},
 \begin{align}\label{eq:matter_effect}
 \sqrt{2}G_{F}N_{e} \simeq 3.8 \times \rho(\text{g/cm}^{3}) \times 10^{-23} \text{ GeV},
 \end{align}
 where $\rho$ is the average Earth matter density along the neutrino path of propagation.
 With $\ldm = 2.55 \times 10^{-3} \text{ eV}^{2}$, the matter effect then turns out to be,
 \begin{align}\label{eq:matter_effect1}
 A = \frac{2\sqrt{2}G_{F}N_{e}E}{\ldm} 
 \simeq 0.03 \times E[\text{GeV}] \times \rho[\text{g/cm}^{3}].
 \end{align}
\begin{table}[t]
\centering
\scalebox{0.94}{
\begin{tabular}{| c | c | c | c | c | c | c | c |}
\hline
$L$ & $\rho$  &
 \multicolumn{2}{| c |}{$\emax{1}$ [GeV]} & 
 \multicolumn{2}{| c |}{$\emax{2}$ [GeV]} & 
\multicolumn{2}{| c |}{$A 
\text{ at } \emax{1}  (\emax{2})$}\\
\cline{3-8}
[km] & $\text{ [gm/cm}^{3}]$ & Approx. & Exact  &
Approx.  & Exact & 
Approx. & Exact \\
&& (Eq.\ \ref{eq:osc_max1}) & (numerically) & 
 &  &&\\
\hline
900 & 2.8 & 1.6 (1.8) & 1.6 (1.8) & 0.6 (0.6) & 0.6 (0.6) & 0.13 (0.05) & 0.13 (0.05)\\
1300 & 2.95 & 2.1 (2.6) & 2.2 (2.6) & 0.8 (0.9)& 0.8 (0.9) & 0.19 (0.07) & 0.20 (0.07)\\
2900 & 3.3 & 3.8 (5.9) & 4.1 (6.1) & 1.7 (2) & 1.8 (2) & 0.37 (0.17) & 0.40 (0.18)\\
\hline
\end{tabular}
}
\caption{\footnotesize{\label{tab:osc_max} This shows the line-averaged densities ($\rho$) estimated using PREM profile~\cite{Dziewonski:1981xy} for the three baselines ($L$); the energies $\emax{1}$ and $\emax{2}$ corresponding to the first and second oscillation maxima in matter (as well as in vacuum , - shown in parentheses); 
and the values of the matter effect term $A$ (see Eq.\ \ref{eq:matter_effect1}) corresponding to  
$\emax{1}$ ($\emax{2}$) for the  
  three neutrino baselines. 
 \textcolor{black}{For $\emax{1,2}$ and $A$, both the approximate values (from Eq.\ \ref{eq:osc_max1}) and 
  the exact numerical values are tabulated for comparison.}}}
\end{table}
 Note that, in writing Eq.\ \ref{eq:prob}, we neglect the {\it{solar}} term 
$\alpha^{2} \sin^{2} 2\theta_{12} c_{23}^{2} \frac{\sin^{2}\hat{A}\Delta_{31}}{\hat{A}^{2}}$ since it is roughly 2 orders of magnitude smaller than the terms shown.
From Eq.\ \ref{eq:prob}, the oscillation maximum for $\pmue$ for a given baseline occurs when the overall factor is maximum (see Appendix \ref{appendix_a} for a detailed justification):
 \begin{align}\label{eq:osc_max}
\sin(1-A)\Delta_{31} \simeq 1
 \implies (1-A)\Delta_{31} =  (2n-1) \frac{\pi}{2}  \text{\qquad}(n = 1, 2, 3..).
 \end{align}
 After properly taking into account the units, the energies corresponding to the oscillation 
 maxima are estimated to be, 
 \begin{align}\label{eq:osc_max1}
  \emax{n}[\text{GeV}] = 1.27 \times \frac{ \ldm[\text{eV}^{2}] \times L [\text{km}]}
 {(2n-1)\frac{\pi}{2} + \frac{G_{F}N_{e}L}{\sqrt{2}}} 
 \simeq 
 \frac{3.2 \times 10^{-3} \times L [\text{km}]}
 {(2n-1)\frac{\pi}{2} + (9.7 \times 10^{-5} \times \rho[\text{g/cm}^{3}] \times L[\text{km}])}.
 \end{align}
$n = 1, 2$ denote the first and second oscillation maxima (OM) respectively. 
The second term ($\propto \rho$) in the denominator of Eq.\ \ref{eq:osc_max1} comes from matter effect. 
Using the average densities ($\rho$) of the three baselines (following the Preliminary Reference Earth Model (PREM)~\cite{Dziewonski:1981xy}~\footnote{\textcolor{black}{Note that PREM profile does not account for local crustal variations due to topography or regional geological structures. However, these local effects are not expected to have any significant impact on the results presented here, and the use of the PREM is therefore adequate for our purposes.}}), the values of $\emax{1}$ and $\emax{2}$ 
are estimated in Table \ref{tab:osc_max}. 
It is evident that, for longer baselines, the oscillation maxima shift to higher energies, thereby making them relatively easier to probe. 
The estimated values of $\emax{1}$ and $\emax{2}$ for 2900 (900) km are approximately 3.8 (1.6) GeV and 1.7 (0.6) GeV respectively. 
This suggests that the EIC-SURF 2900 km baseline could be a very promising option, as it allows the possibility of observing more than one oscillation cycle. 
Furthermore, the presence of the overall factor $(1-A)$ in the denominator of Eq.\ \ref{eq:prob} 
increases the amplitude of oscillation with energy and matter density, - which in turn increases with baseline (see Table \ref{tab:osc_max} for an estimate of the factor $A$ at first and second oscillation maxima). 
Thus the magnitude of $\pmue$ at 2900 km is almost twice than at 900 km around the first oscillation maximum (as evident from Fig.\ \ref{fig:flux_xsec}: top row, 3rd column). 

 In vacuum ($\rho, A \to 0$) Eq.\ \ref{eq:prob} reduces to,
 \begin{align}\label{eq:prob_vac}
 P_{\mu e}^{\text{vac}} &\simeq 
  \sin\Delta_{31} \bigg[\sin^{2}2\theta_{13}\sin^{2}\theta_{23}
 \sin\Delta_{31} 
 +
  \alpha  \sin2\theta_{13} \sin2\theta_{23}  \sin 2\theta_{12}  
\Delta_{31}\cos (\da + \Delta_{31})  \bigg].
 \end{align}
Also from Eq.\ \ref{eq:osc_max1} it is evident that the oscillation maxima in vacuum shift to higher energies compared to that in matter, with $\emax{2} = \emax{1}/3$. 
The influence of matter effects grows with the baseline length ($L$) and becomes increasingly pronounced at higher energies, as reflected in the deviations of $\emax{1,2}$ from its vacuum value. 
Although both $\emax{1}$ and $\emax{2}$ are shifted, the relative change is larger for $\emax{1}$. These shifts are most significant for the longest baseline considered ($L = 2900$ km).

 \begin{figure}[b]
 \centering
 \includegraphics[scale=0.55]{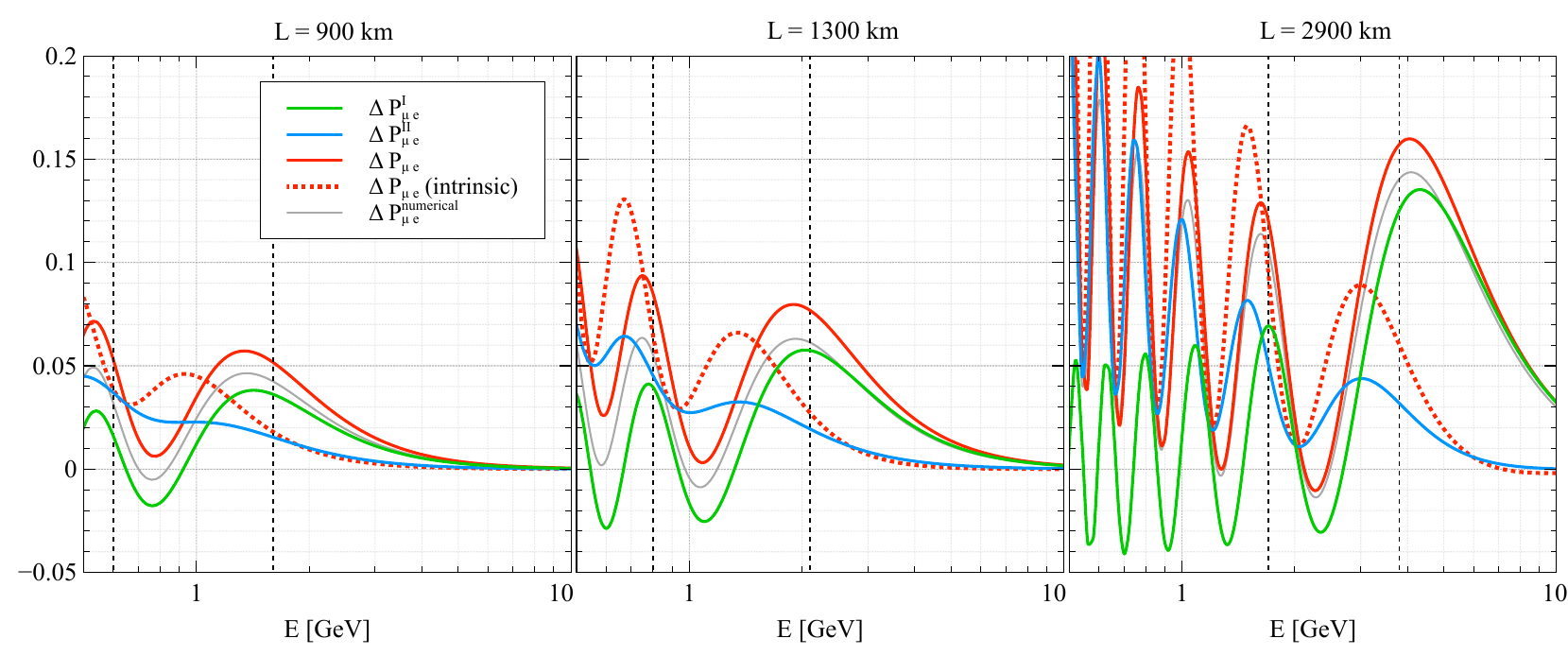}
 \caption{\footnotesize{The solid curves show the quantities $\acp^{\text{I}}, \acp^{\text{II}}$ and 
 $\acp = (\acp^{\text{I}} + \acp^{\text{II}})$ from Eq.\ \ref{eq:acp} as functions of energy 
 for the three baselines (900 km, 1300 km, 2900 km). 
 The dotted red curves indicate the intrinsic CP asymmetry $(\acp - \acp^{\da=0})$ 
 originating only from the phase $\da$ and free from matter effects.
 \textcolor{black}{For comparison, we have also included the exact values of $\acp$ calculated 
 numerically using GLoBES.}
 The pair of black dotted vertical lines in each panel indicates 
 the relevant $\emax{1}$ and $\emax{2}$.}}
  \label{fig:asym_analysis}
 \end{figure}

In order to understand CP violation, we now proceed to calculate the CP asymmetry $\acp = \pmue - \pmuebar$ (using Eq.\ \ref{eq:prob}). 
For $\muebar$ oscillation, the sign of the matter effect and the CP phase gets changed, \ie, 
 $A \to -A$ and $\da \to -\da$. Thus,
\begin{align}\label{eq:acp}
&\acp 
\simeq 
\underbrace{\sin^{2} \tb\,\sin^{2}\tc 
\bigg[ 
\frac{\sin^{2}(1-A)\Delta_{31}}{(1-A)^{2}} 
- \frac{\sin^{2}(1+A)\Delta_{31}}{(1+A)^{2}} 
\bigg]}_{\acp^{\text{I}}} \nonumber \\
&+ 
\underbrace{\alpha\sin 2\ta \sin 2\tb \sin 2\tc \frac{\sin A\Delta_{31}}{A}
\bigg[
\frac{\sin(1-A)\Delta_{31}}{1-A}\cos(\da + \Delta_{31}) 
  -\frac{\sin(1+A)\Delta_{31}}{1+A}\cos(\da - \Delta_{31})
\bigg].}_{\acp^{\text{II}}}
\end{align}
The term $\acp^{\text{I}}$ in the right hand side originates purely due to matter effect $A$ 
and its amplitude increases with energy. 
The term $\acp^{\text{II}}$ is a function of the CP phase $\da$ and also has some matter dependence. Note that, in vacuum, $A \to 0$ such that $\acp^{\text{I}}$ vanishes and the asymmetry simplifies to, 
\begin{align}\label{eq:acp_vac}
\acp \Big |_{A \to 0} \simeq -2\alpha \sin 2\ta \sin 2\tb \sin 2\tc 
\Delta_{31} \sin^{2} \Delta_{31} \sin \da. 
\end{align}
The amplitude of $\acp^{\text{II}}$ tends to increase with decreasing energy (due to  the overall factor $1/A$). 
Importantly, this implies that the pure CP violation sensitivity arising from the phase $\da$ improves as higher oscillation maxima are probed for a given baseline. 
At 2900 km, the more accessible values of $\emax{1}$ and $\emax{2}$ therefore make the EIC-SURF baseline a particularly promising option for investigating leptonic CP violation. 

 \begin{figure}[b]
 \centering
 \includegraphics[scale=0.5]{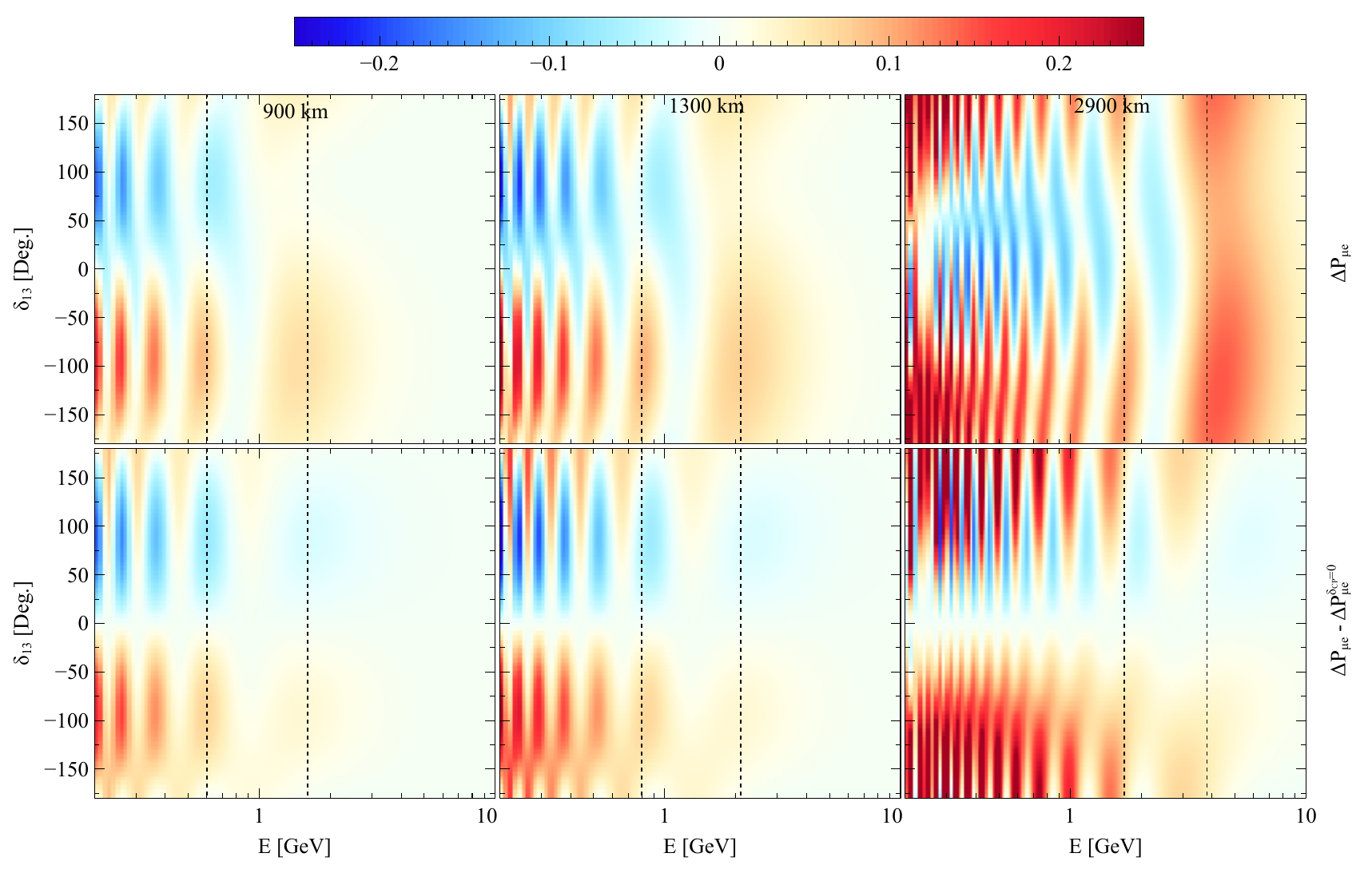}
 \caption{\footnotesize{The top row shows the heatplots of the CP asymmetry $\acp = (\pmue -\pmuebar)$ 
 for the three baselines (900 km, 1300 km, 2900 km) in the parameter space of the CP phase $\da$ and energy E. 
 The bottom row shows the heatplots of the intrinsic CP asymmetry $(\acp - \acp^{\da=0})$. 
 The pair of black dotted vertical lines in each panel indicates 
 the relevant $\emax{1}$ and $\emax{2}$.}}
  \label{fig:asym_heatplot}
 \end{figure}

In this context, it is worth noting that for large baselines (\eg, 2900 km), matter effects become significant and must be carefully disentangled from genuine CP violation effects~\cite{Arafune:1997hd, Nunokawa:2007qh}. 
In such cases, a more appropriate way to quantify the intrinsic CP asymmetry arising from the phase $\da$ is to subtract the asymmetry induced purely by matter effects in the absence of the CP violating phase $\da$. 
This is expressed as $(\acp - \acp^{\da=0})$~\cite{Ohlsson:2013ip, Rout:2017udo}.
 
In Fig.\ \ref{fig:asym_analysis} we show how the two contributing terms in Eq.\ \ref{eq:acp}, $\acp^{\text{I}}$ and
$\acp^{\text{II}}$, as well as the whole quantity $\acp = (\acp^{\text{I}} + \acp^{\text{II}})$ change with energy for the three baselines. 
We keep the oscillation parameters including the CP violating phase $\da$ fixed to their best fit values (Table \ref{tab:parameters}).
We also show the intrinsic CP asymmetry $(\acp - \acp^{\da=0})$ as discussed above. 
The frequency of oscillation increases with baseline and for a given baseline, decreases with energy (due to the presence of the term $\Delta_{31}$ in the arguments of the sine and cosine terms in Eq.\ \ref{eq:acp}). 
The interplay between $\acp^{\text{I}}$ and $\acp^{\text{II}}$ offers a relatively higher 
CP asymmetry for the 2900 km baseline. 
Clearly, the intrinsic CP asymmetry which follows a spectral pattern similar to $\acp^{\text{II}}$  is much higher at for the 2900 km than the other two baselines.

In order to analyse the CP asymmetries when $\da$ is varied, we 
show the heatplots of $\acp$ and $(\acp-\acp^{\da=0})$ (both calculated 
numerically using GLoBES) in the 
plane of $E-\da$ in Fig.\ \ref{fig:asym_heatplot} for the three 
baselines. 
The red (blue) colour indicates a positive (negative) value while the  saturation of the colour signifies the magnitude 
of the asymmetry. 
As expected, the magnitude of the asymmetry is highest around the 
maximally CP violating values $\da \simeq \pm \pi/2$. 
For simplicity, the behaviour of $\acp$ for small matter effects can be  
understood from the vacuum approximation (Eq.\ \ref{eq:acp_vac}).  
The $\sin\da$ factor with a minus sign in Eq.\ \ref{eq:acp_vac} 
explains why $\acp$ is negative (positive) for $\da > 0$ ($\da < 0$). 
Such antisymmetry around $\da \simeq 0$ gets distorted by 
matter effects,- most prominently for 2900 km as illustrated in Fig.\ \ref{fig:asym_heatplot}. 
The bottom row of Fig.\ \ref{fig:asym_heatplot} illustrates 
the intrinsic asymmetries largely free from matter effects and with 
less distorted patterns.

The heatmaps clearly show that as the baseline increases, multiple oscillation cycles can, in principle, be probed. This, in turn, leads to enhanced CP asymmetry across the accessible energy range. 
To get a simple numerical estimate of the number of oscillation cycles 
present within a given energy interval for a fixed $\da$, we use the vacuum approximation 
(Eq.\ \ref{eq:acp_vac}) for brevity. 
The number of cycles is determined by the argument of the term $\sin^{2}\Delta_{31} = (1 - \cos2\Delta_{31})/2$. 
Since $2\Delta_{31} \simeq 2\pi \times 10^{-3} \times L[\text{km}]/E[\text{GeV}]$, the 
number of cycles $N_{\text{cycles}}$ in the energy interval $[E_1,E_2]$ (in GeVs) is given by,
\begin{align}\label{eq:osc_cycles}
N_{\text{cycles}} \simeq 10^{-3} \times L[\text{km}] 
\times \bigg(\frac{1}{E_1}- \frac{1}{E_2}\bigg).
\end{align}
Thus in the range $[0.2-10]$ GeV, the estimated number of oscillation cycles 
for the three baselines (900, 1300, 2900 km) are 4.4, 6.4, 14.2 
respectively. 
This approximate analytical estimate is in good agreement with the  numerical simulation, as reflected by the number of red or blue {\it{blobs}} observed in each panel of Fig.\ \ref{fig:asym_heatplot}. 
\textcolor{black}{
In this context, it needs to be mentioned that counting the number of oscillation 
cycles in actual data using Eq.\ \ref{eq:osc_cycles} requires both very good energy resolution (especially at lower energies), as well as small uncertainties in the neutrino flux (\ie, less systematics). 
Nevertheless, Eq.\ \ref{eq:osc_cycles} can potentially give an instant, semi-quantitative insight into the 
oscillatory nature of $\mue$ transition for a given baseline and energy interval. 
}
\begin{figure}[b]
 \centering
 \includegraphics[scale=0.6]{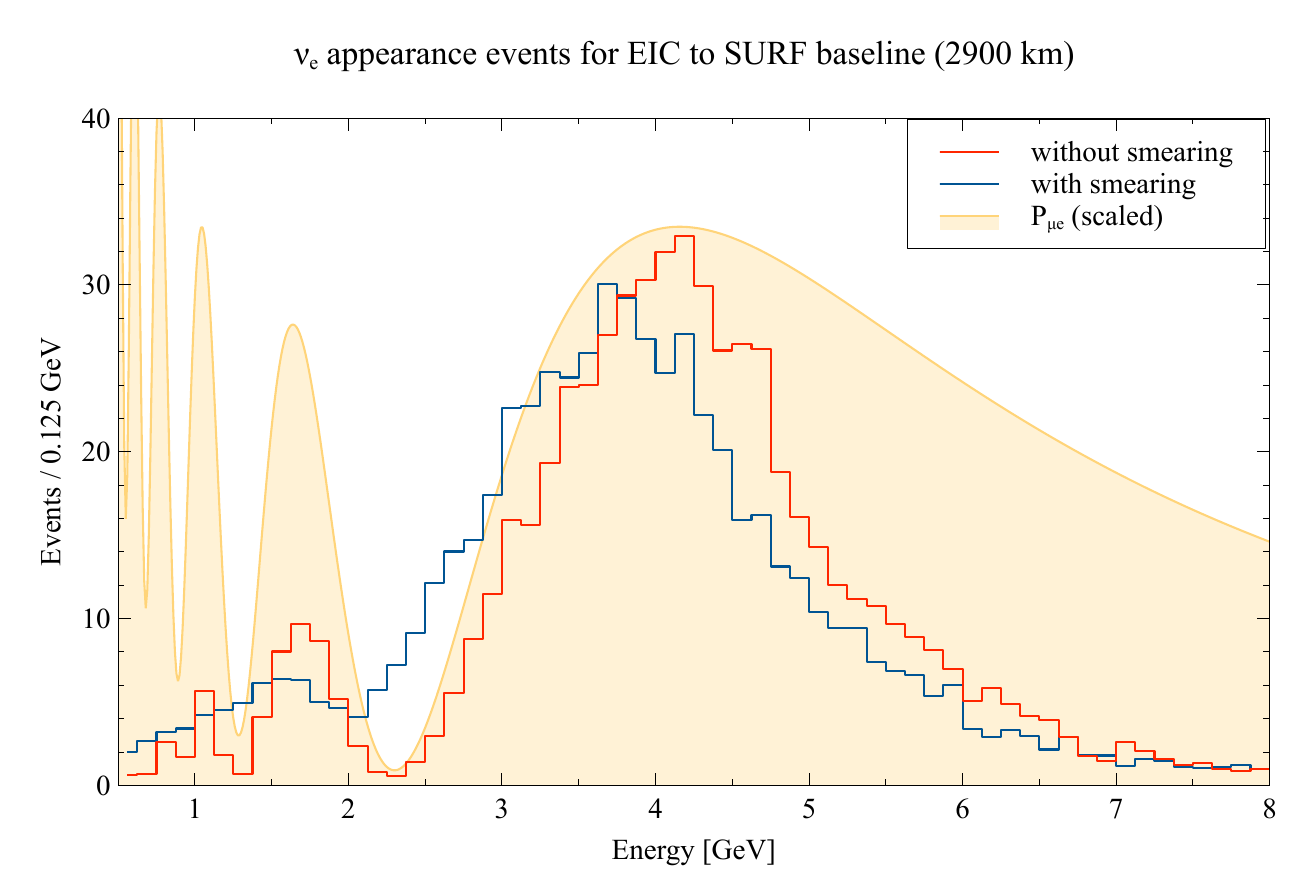}
 \caption{\footnotesize{$\nue$ appearance events (signals only) at a 17 kt WbLS detector for the EIC to SURF baseline (2900 km). The blue curve indicates events for a WbLS detector with a realistic energy smearing, while the red curves show the case of no energy smearing. The light yellow regions show the underlying $\pmue$ (scaled with a constant arbitrary factor for better visibility).}}
  \label{fig:event_nue_2900}
 \end{figure}
\section{Event spectra}
\label{sec:event}
Using the neutrino flux at EIC we generate event spectra at the detectors at SURF (2900 km baseline) and at SNOLAB (900 km baseline) sites. 
For 2900 km, we use the neutrino flux simulated with a horn current of 
493 kA, while for the 900 km baseline the flux generated with a 93 kA horn current is used. 
 \begin{figure}[b]
 \centering
 \includegraphics[scale=0.6]{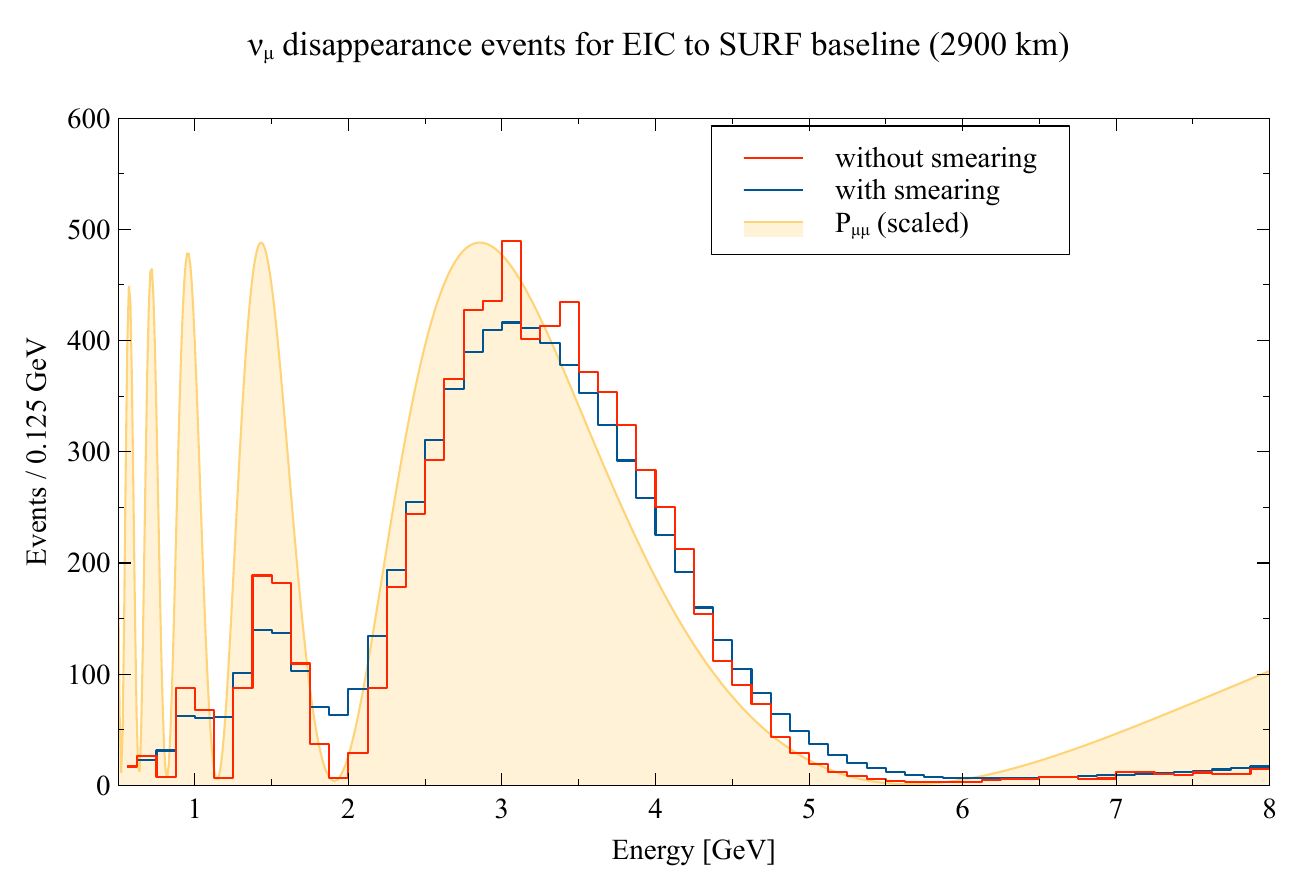}
 \caption{\footnotesize{$\numu$ disappearance events (signals only) at a 17 kt WbLS detector for the EIC to SURF baseline (2900 km). The blue curve indicates events for a WbLS detector with a realistic energy smearing, while the red curves show the case of no energy smearing. The light yellow regions show the underlying $\pmue$ (scaled).}}
  \label{fig:event_numu_2900}
 \end{figure}
We assume a 7 yrs. of runtime (3.5 yrs. each in neutrino and antineutrino mode) and a water based liquid scintillator or WbLS~\footnote{WbLS can detect particles based on both scintillation and Cherenkov radiation, the latter providing a directional capability as well~\cite{OrebiGann:2015gus, Fischer:2018zsr, Kaptanoglu:2019gtg, Callaghan:2023oyu}. 
WbLS is a proposed detector in the THEIA collaboration~\cite{Theia:2019non, Theia:2022uyh, mary_dune_theia_2023}. WbLS detectors can help in CP violation studies~\cite{Ge:2022iac}; in antineutrino detection~\cite{Zsoldos:2022mre}; in low energy astrophysical neutrino detection~\cite{DeGouvea:2020ang, Sawatzki:2020mpb, DeRomeri:2021xgy, Borboruah:2025hai}; in neutrino-nucleon interaction studies~\cite{ANNIE:2023yny}; as well as in studying rare processes such as neutrino decay or absolute neutrino mass~\cite{DeGouvea:2020ang, Parker:2023cos, Chauhan:2022wgj}.}  
with a fiducial mass of 17 kt for both the baselines. 
Following the discussions in \cite{Theia:2019non}, 
regarding the capability of WbLS to identify Cherenkov rings, we implement separate 
{\it{channels}} in the corresponding {\it{.glb}} file of GLoBES for $\nue$ signals: 
one, two or three-ring {\it{rules}} with either zero or one Michel electrons from stopped pion or muon decay. 
The channels with $n$ Cherenkov rings and $i$ Michel electrons are referred to as $nRiD$ with $n=1,2,3$ and $i=0,1$ in this case. 
We further implement different {\it{channels}} based on whether the $\nue$ signal is charge-current quasi-elastic (CCQE) or charge-current non-quasielastic (CCNQE). 

For the $\numu$ signals we implement two separate {\it{channels}}, namely $1R0D, 1R1D$  and consider charge current interactions inclusively (CC), \ie, both CCQE and CCNQE together in one {\it{channel}} in GLoBES. 
The single ring channels provide the dominant contributions to the $\nue$ and $\numu$ 
event spectra. 
All these {\it{channels}} include pre and post smearing efficiencies taken from \cite{guang_theia_2018, Theia:2019non}. 
For instance, in the energy range of 1-5 GeV, the pre (post) smearing efficiencies for the {\it{channel}} $1R0D$ for CCQE $\nue$ signal lie around $75\%-85\%$ ($85\%-95\%$).  
For $\numu$ channels, we have used an overall flat efficiency of $90\%$ 
throughout the entire energy range. 
For the same $1R0D$ {\it{channel}}, the migration matrices for smearing in the WbLS detector correspond to an energy resolution ($\sigma/E$) of $6\%-15\%$ for CCQE $\nue$ signal in 1-5 GeV energy range. 
For the $\numu$ channel, we have assumed a Gaussian smearing of $10\%$. 
In order to understand the impact of smearing, we have also 
estimated the event spectra by assuming a detector with perfect energy 
resolution ($\ie$, no energy smearing).

 \begin{figure}[t]
 \centering
 \includegraphics[scale=0.6]{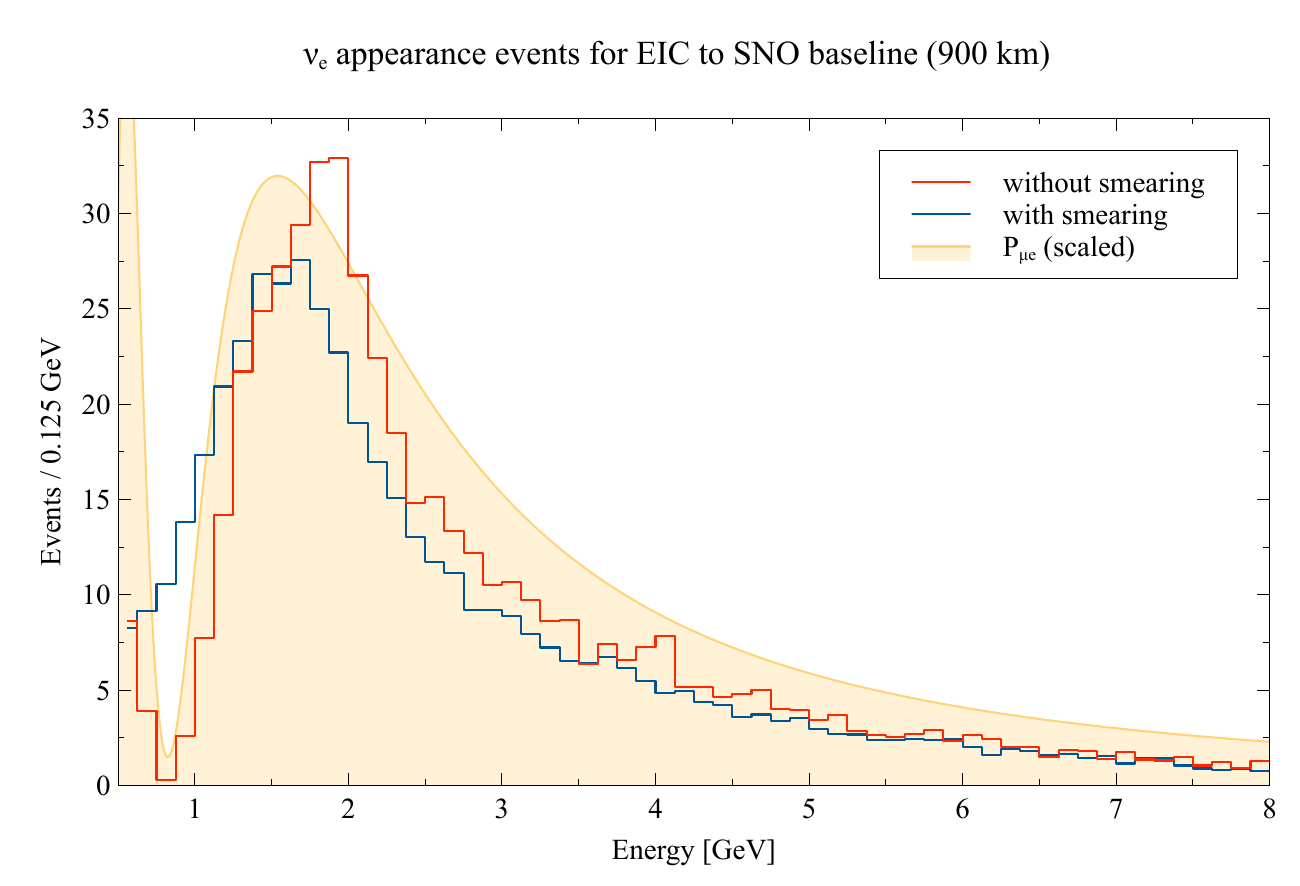}
 \caption{\footnotesize{$\nue$ appearance events (signals only) at WbLS detector for the EIC to SNOlab baseline (900 km).}}
  \label{fig:event_nue_900}
 \end{figure}

To better illustrate how the two baselines trace the oscillation cycles, we focus exclusively on the signal contributions from the $\mue$ and $\mumu$ channels, omitting backgrounds for simplicity.
In Figs.\ \ref{fig:event_nue_2900} and \ref{fig:event_numu_2900} we show the $\mue$ and $\mumu$ signals respectively (both with 
and without energy smearing) at the WbLS detector placed at SURF. 
The spectra are superimposed with the underlying $\pmue$ (scaled with an 
arbitrary factor for better visualization). 
The smeared $\mue$ appearance spectra can track the underlying oscillation upto around 5 GeV, showing both 
the first and second oscillation maxima clearly. 
For a WbLS detector with near-perfect energy resolution, upto three 
oscillation maxima in the $\mue$ channel can clearly be observed. 
At energies $\gtrsim$ 5 GeV, the rapidly falling flux also diminishes 
the events. 
The $\mumu$ disappearance spectra in Fig.\ \ref{fig:event_numu_2900} also tracks the underlying $\pmumu$ oscillation patterns quite faithfully for both with 
and without energy smearing. 

\begin{figure}[t]
 \centering
 \includegraphics[scale=0.6]{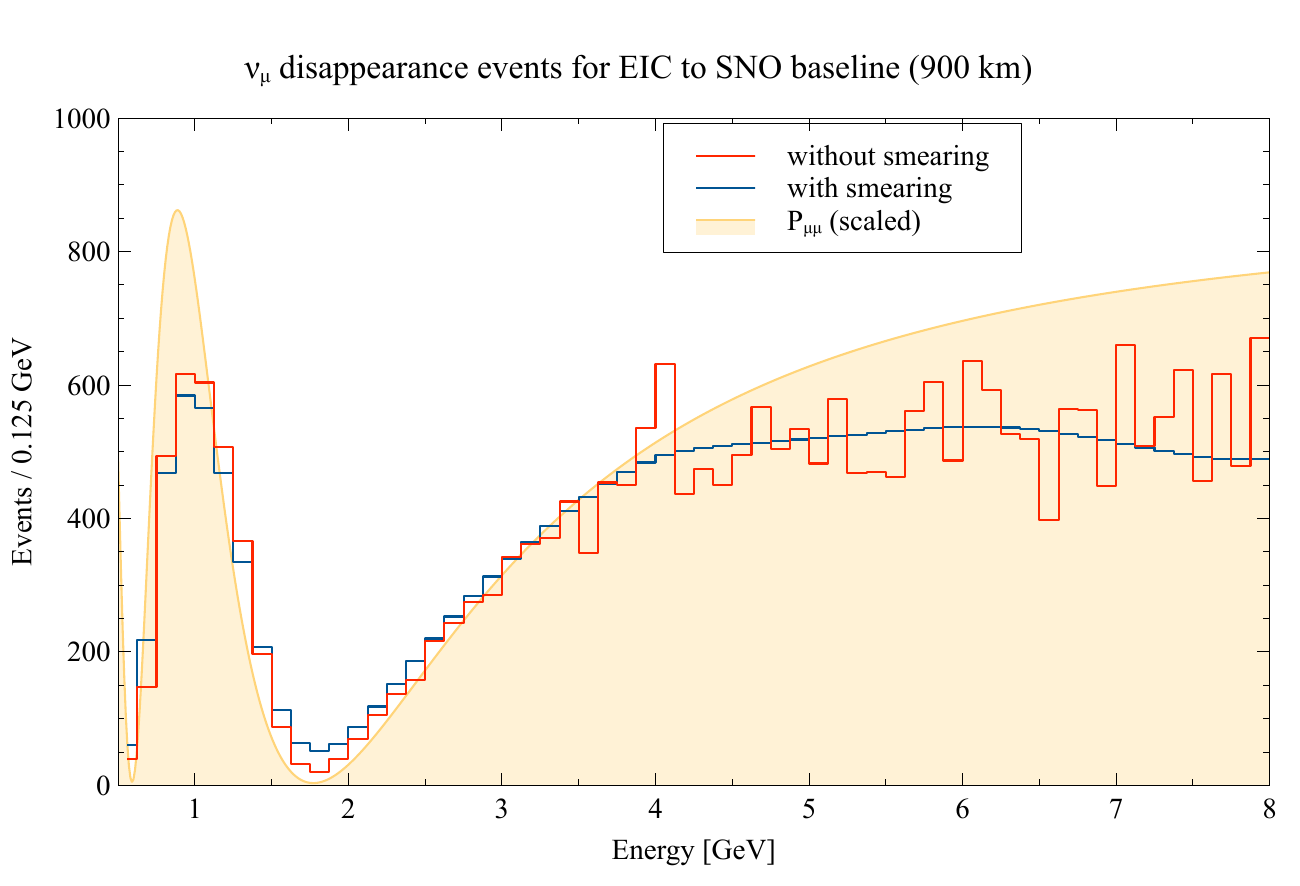}
 \caption{\footnotesize{$\numu$ disappearance events (signals only) at WbLS detector for the EIC to SNOlab baseline (900 km).}}
  \label{fig:event_numu_900}
 \end{figure}

Similarly, in Figs.\ \ref{fig:event_nue_900} and \ref{fig:event_numu_900} we show 
the event spectra for the $\mue$ and $\mumu$ channel respectively. 
The spectra follow the probability patterns faithfully but can reconstruct 
only one oscillation cycle (first oscillation maximum) confidently, - even with a near-perfect energy response. 
The number of events around the peak of the $\mue$ spectra for 900 km 
is at the similar ballpark as that for 2900 km, - an observation 
consistent with Fig.\ \ref{fig:flux_xsec} (compare the solid red and solid blue curves in the bottom-left panel). 
\section{$\chisq$ Analysis}
\label{sec:chisq}
Finally we estimate the $\chisq$ sensitivities to CP violation for the WbLS 
detectors at SURF and at SNO. 
Below we describe the analytical form of $\chisq$. 
\begin{align}\label{eq:chisq}
\Delta \chi^{2}(\da) &= {\text{Min}} 
\bigg[
\color{black}
\underbrace{\color{black}
2\sum_{x}^{\text{mode}}\sum_{j}^{\text{channel}}\sum_{y}^{\text{bin}}
\bigg\{
N_{yjx}^{\text{fit}}(\da^{\text{fit}}=0,\pi) - N_{yjx}^{\text{data}}(\da) 
+ N_{yjx}^{\text{data}}(\da) \ln\frac{N_{yjx}^{\text{data}}(\da)}{N_{yjx}^{\text{fit}}(\da^{\text{fit}}=0,\pi)} 
}_{\text{statistical}}
\bigg\} 
\color{black}
\nonumber \\
&+ 
\color{black}
\underbrace{\color{black}
\sum_{l}\frac{(p_{l}-p^{\text{fit}}_{l})^{2}}{\sigma_{p_{l}}^{2}}
}_{\text{prior}}
\color{black}
+ 
\color{black}
\underbrace{\color{black}
\sum_{k}\frac{\eta_{k}^{2}}{\sigma_{k}^{2}}
}_{\text{systematics}}
\color{black}
\bigg],
\end{align}
where 
the index $y$ is summed over the energy bins in the range $0.5-18$ GeV\,\footnote{We have a total of $65$ energy bins in the range $0.5-18$ GeV: $60$ bins each having a width of 
$0.125$ GeV in the energy range of $0.5 - 8$ GeV and $5$ bins
with 2 GeV width in 8-18 GeV. 
Also note that, the analysis 
range for the SURF (SNO) site was taken as $0.8-18$ $(2.2-18)$ GeV for the case where only the first oscillation maximum was consider by ignoring the second.} unless otherwise mentioned explicitly. 
The index $j$ corresponds to the six appearance channels $nRiD$ 
($n$ Cerenkov rings and $i$ Michel electrons from stopped pion or muon decay) with $n=1,2,3$, and $i=0,1$, as discussed in Sec.\ \ref{sec:event}. 
Finally the index $x$ runs over the neutrino and antineutrino modes.
$N^{\text{data}}$ corresponds to the set of events generated for any value of the CP phase $\da \in [-\pi,\pi]$ with the other oscillation parameters fixed to the global best fit values.
 $N^{\text{fit}}$ corresponds to the events generated only for the CP 
 conserving values of the phase: $\da^{\text{fit}}=0,\pi$ with other oscillation parameters being varied in their allowed range (see Table\ \ref{tab:parameters} for both the best fit values and the range of variation of the oscillation parameters). 
The terms in the first row of the right-hand-side of Eq.~\eqref{eq:chisq} correspond to the statistical contribution. 

The first two terms in the statistical contribution to the $\chisq$ constitute the algebraic difference 
($N^{\text{fit}} - N^{\text{data}}$) while \textcolor{black}{the logarithmic-term proportional to 
$\ln(N^{\text{data}}/N^{\text{fit}})$ takes into account the relative difference or ratio of the two sets of events}%
\footnote{
Note that the definition of $\chisq$ described in Eq.\ \ref{eq:chisq} is Poissonian in nature. In the limit of large number of events (N),  this reduces to the Gaussian form : 
\begin{align*}
\lim_{N \to \infty} \Delta \chi^{2}(\da)  \simeq {\text{min}}
\Bigg[\sum_{x}^{\text{mode}}\sum_{j}^{\text{channel}}\sum_{i}^{\text{bin}}
\frac{
\Big\{
N_{ijx}^{\text{fit}}(\da=0,\pi) - N_{ijx}^{\text{data}}(\da)
\Big\}^{2}}
{N_{ijx}^{\text{data}}(\da)}
+ \text{prior} + \text{systematics}\Bigg].
\end{align*}}. %

The two terms in the second line of Eq.\ \ref{eq:chisq} correspond to the prior and systematics respectively. 
The \textit{prior} term accounts for the penalty of the $l$ 
number of \textit{fit} parameters deviating away from the 
corresponding $p^{\text{data}}$. 
 \begin{figure}[t]
 \centering
 \includegraphics[scale=0.6]{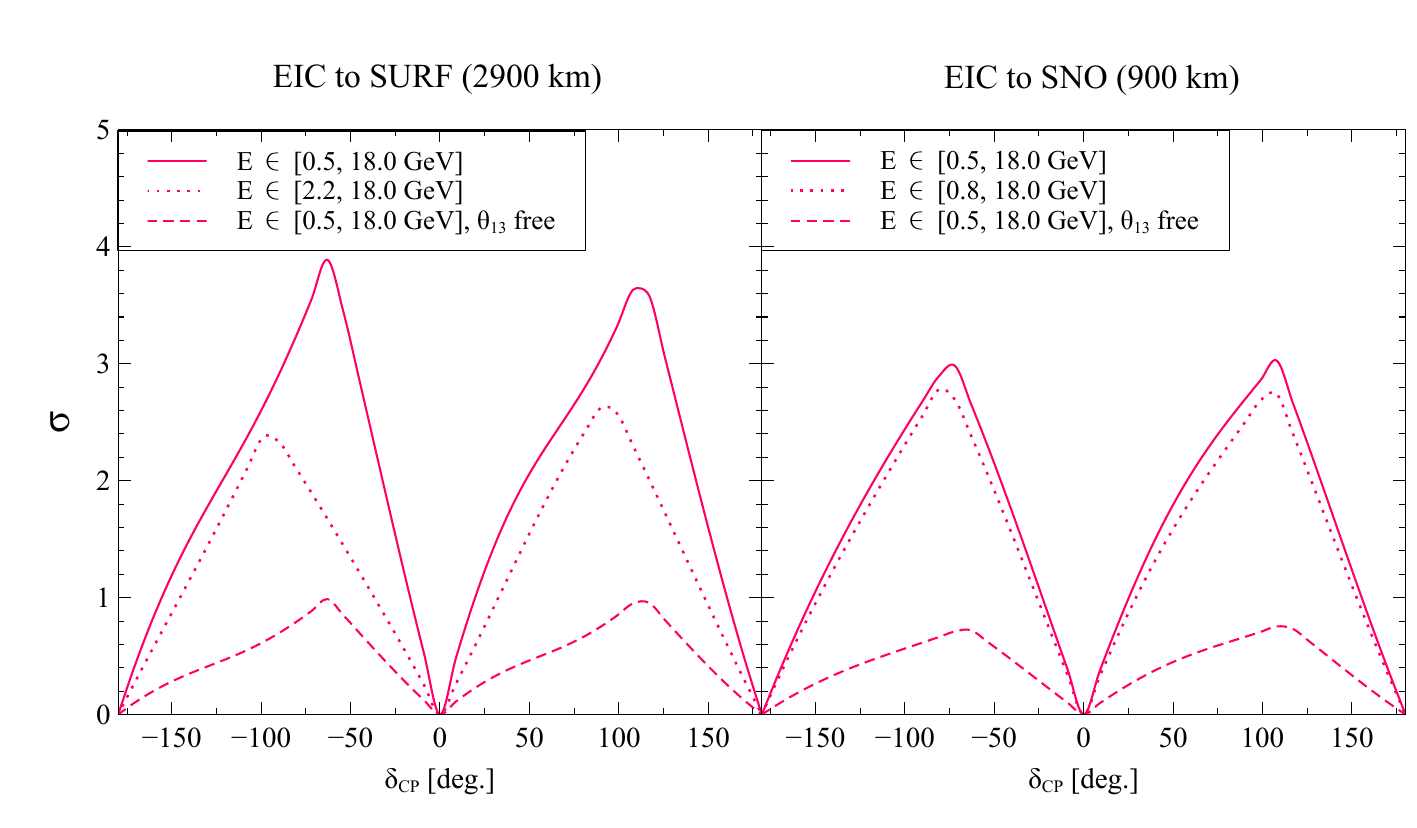}
 \caption{\footnotesize{This shows the CP violation sensitivities achieved from the neutrino beam (produced using 1 MW proton beam at BNL) at a 17 kt WbLS detector at SURF (left panel), and at SNO (right panel) site for a $(3.5+3.5)$ yrs. of $(\nu + \bar{\nu})$ mode runtimes.  
 The sensitivities shown by the dotted curves were estimated using a curtailed energy range so as to consider only the first oscillation maximum, while the solid and dashed curves additionally consider second and higher oscillation maxima as well.}}
  \label{fig:chisq_cpv}
 \end{figure}
The degree of this deviation is controlled by  $\sigma_{pl}$ which is the uncertainty in 
the prior measurement of the best-fit values of $p^{\text{data}}$ (see the last column of Table \ref{tab:parameters} 
for the values used in the present analysis). 
The  \textit{systematics}-term accounts for the variation of the systematic/nuisance parameters.
$\eta$ is the set of values of $k$-systematics parameters $\{\eta_{1}$, $\eta_{2}$, $\dots$ $\eta_{k}\}$ (which also impact $N^{\text{fit}}$ as normalization uncertainties)  while 
$\sigma_k$ is the uncertainty in the corresponding  systematics. 
 This way of treating the nuisance parameters in the $\chisq$ calculation is known as the {\it{method of pulls}}~\cite{Huber:2002mx,Fogli:2002pt,GonzalezGarcia:2004wg,Gandhi:2007td}. 
Regarding the systematics~\cite{Theia:2019non}, the $\nu_{e}$ and $\bar{\nu}_{e}$ signal modes have a normalization uncertainty of $2\%$ each. 
As discussed in Sec.\ \ref{sec:event}, backgrounds were neglected in the 
$\chisq$ analysis.

The final estimate of the $\chisq$ as a function of the CP phase $\da$ is obtained 
after minimizing the entire quantity within the square bracket in Eq.\ \ref{eq:chisq} over the CP conserving values of the phase ($\da^{\text{fit}} = 0, \pi$); the {\it{fit}} oscillation parameters in their allowed range; as well as over the systematics $\eta$. 
This minimization is also referred to as marginalization.
Technically, this  procedure is the frequentist method of hypotheses testing~\cite{Fogli:2002pt, Qian:2012zn}.

 \begin{figure}[t]
 \centering
 \includegraphics[scale=0.6]{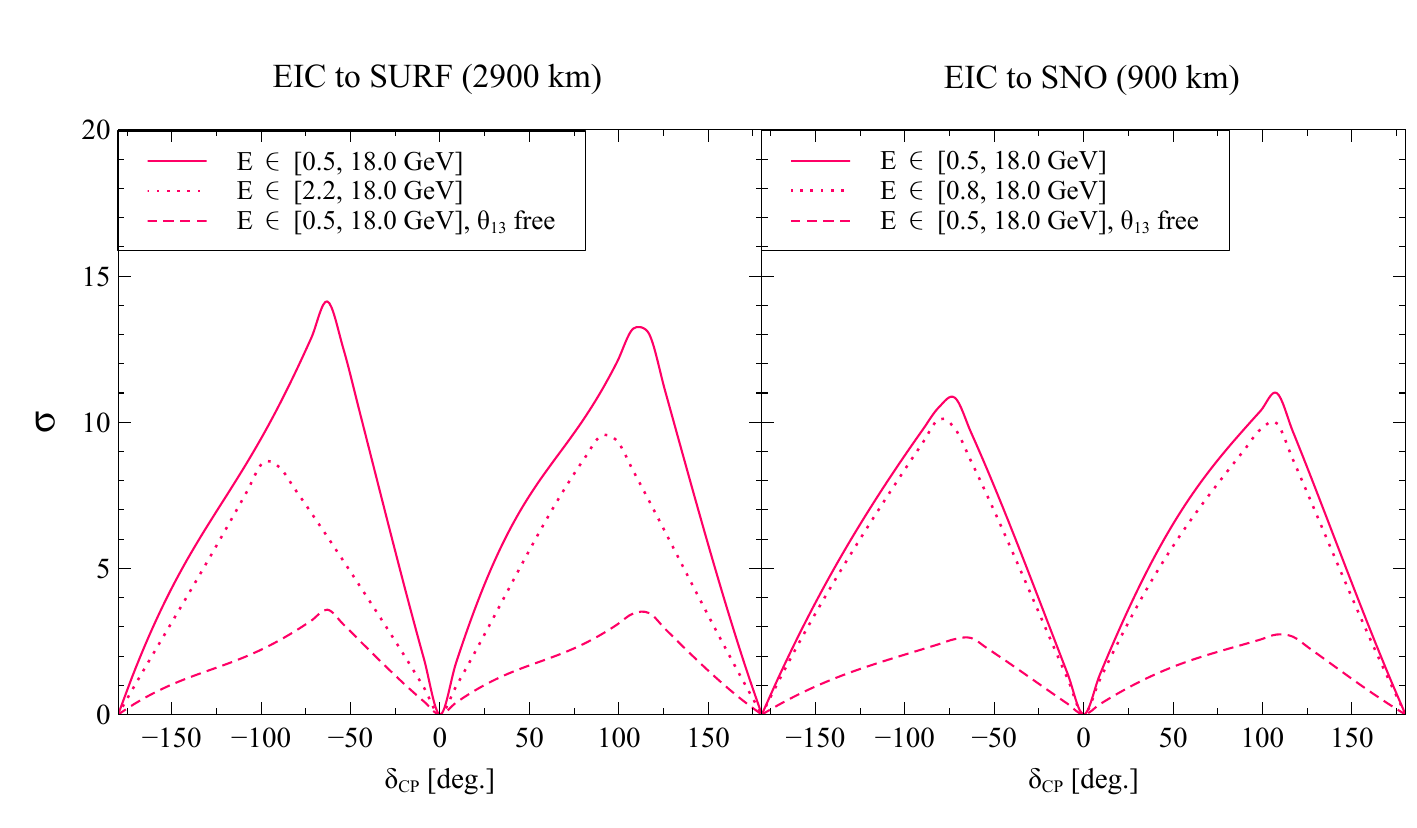}
 \caption{\footnotesize{Similar to Fig.\ \ref{fig:chisq_cpv}, but using the full proton beam power of 13.2 MW.}}
  \label{fig:chisq_cpv_high_mw}
 \end{figure}

In Fig.\ \ref{fig:chisq_cpv} we show the CP violation sensitivities ($\sigma = \sqrt \chisq$) of the WbLS 
detectors at SURF and SNO site for a total runtime of 7 yrs. (3.5 yrs. each in $\nu$ and $\bar{\nu}$ modes). 
The solid curves show the case when we take the full energy interval 
($0.5-18$ GeV) for analysis which considers more than one oscillation maxima. 
With a realistic energy resolution the WbLS detectors can achieve sensitivities above $3.5\sigma$ (SURF), and almost $3\sigma$ (SNO). 
The dotted curves show the sensitivities with a narrower energy range ($2.2-18$ GeV for SURF and $0.6-18$ GeV for SNO) in order to consider only the first oscillation maximum in the analysis. 
Using only first oscillation maximum significantly reduces the sensitivities for SURF (Around $3\sigma$ for perfect energy reconstruction and around $2.5 \sigma$ with a finite energy smearing) as compared to SNO (around $3.5\sigma$ for no smearing and slightly below $3\sigma$ with a realistic energy smearing). 
This demonstrates the crucial role played by the second (and higher) oscillation maxima in probing CP violation for longer baselines. 
 
So far, in generating both the solid and the dotted curves, we have    
marginalised over the less-well known oscillation parameters $\tc$ and $\ldm$ with some prior uncertainties (see Table \ref{tab:parameters}), \textcolor{black}{while keeping the rest of the parameters ($\ta, \tb, \sdm$) fixed to their best fit values (also listed in Table \ref{tab:parameters}).} 
Finally, to complete our study we also show the conservative scenario when $\tb$ is also 
\textcolor{black}{marginalized over (in addition to the marginalization of $\tc$ and $\ldm$),} - which brings down the CP sensitivities to around $1\sigma$ (dotted curves). 
Note that, for the dotted curves we have taken the full energy range into account in order to consider multiple oscillation maxima. 

In Fig.\ \ref{fig:chisq_cpv_high_mw} we show the similar $\chisq$ results with the assumption that the beam has the full power of 13.2 MW instead of 1 MW. 
In this optimistic case, the CP sensitivities can be very high, - well above $5\sigma$ even with using only the first oscillation maximum. 

\section{Summary and Conclusion}
\label{sec:summary}
The Electron-Ion Collider (EIC) is an upcoming next-generation particle accelerator designed primarily to study nuclear structure with unprecedented precision by colliding high-energy electron beams with high-energy proton and ion beams.
In this work, we have explored how this powerful proton beam can additionally be utilized to generate intense neutrino beams, which can then be studied for oscillation effects using two water-based liquid scintillator (WbLS) detectors,-one located at SURF (with a baseline of 2900 km) and another at the SNO site (900 km baseline).

The proton beam to be used in EIC will be both highly polarized and highly energetic (275 GeV), the first of its kind worldwide. Using Geant4 simulations, we model a 1 MW fraction of the EIC proton beam directed onto a graphite target to produce secondary hadrons. 
These hadrons are subsequently allowed to decay, generating a wide-band neutrino flux (with energies up to several GeV), after being focused by magnetic horns operating at currents between 93 kA and 593 kA.
We find that horn currents of 93 kA and 493 kA yield neutrino fluxes optimally suited to enhance the $\nu_\mu \rightarrow \nu_e$ oscillation signatures for the 900 km and 2900 km baselines, respectively.

We illustrate the optimized neutrino fluxes at both baselines, along with the corresponding $\pmue$ probabilities, neutrino cross sections, and their products, to provide an overview of the expected event spectra. 
We then discuss the analytical behaviour of $\pmue$ at different baselines, highlighting the emergence of multiple oscillation maxima.
For longer baselines, both the first and second oscillation maxima shift toward higher energies, making them more experimentally accessible. 
For the 2900 km baseline, the first and second maxima appear around 3.8 GeV and 1.7 GeV, respectively,- energies that lie well within the reach of the EIC-produced neutrino beam.
Crucially, probing multiple oscillation maxima significantly enhances the sensitivity to leptonic CP violation. 
Using analytical expressions and heatmap visualizations, we demonstrate how CP asymmetries vary with $\delta_{\text{CP}}$ across different baselines.

Using GLoBES, we simulate neutrino event spectra for both the $\nu_e$ appearance and $\nu_\mu$ disappearance channels in a 17 kt (fiducial) WbLS detector placed at SURF (2900 km) and SNO (900 km).
We show that the simulated events, both with ideal energy reconstruction and realistic smearing, faithfully reproduce the oscillation features of the underlying probabilities.
Finally, we perform a $\chisq$ analysis to evaluate the CP-violation sensitivities at both baselines.
Our results demonstrate the critical role of the second oscillation maximum, particularly for longer baselines in enhancing CP violation discovery potential.
For example, with realistic energy smearing, the WbLS detector at SURF (2900 km) achieves a maximum CP sensitivity of about $3.5\sigma$ or higher, while excluding the second oscillation maximum reduces this sensitivity to around $2.5\sigma$. 

\textcolor{black}{It should be noted that this study uses simplified systematics and neglects backgrounds; a full simulation including these effects is reserved for future work.}

In conclusion, our study highlights that a fraction of the high-energy, polarized proton beam from the EIC can be effectively utilized to produce a GeV-scale neutrino beam, which can be detected at existing underground facilities to explore key aspects of neutrino oscillation physics.
Such an experimental realization would enable the simultaneous measurement of multiple oscillation maxima-an achievement that remains challenging for current long baseline experiments, and thereby provide a powerful probe of leptonic CP violation.

\section*{Appendix}
\appendix
\renewcommand{\theequation}{\thesection.\arabic{equation}}
\setcounter{equation}{0}
\renewcommand{\thesection}{\Alph{section}}
\section{Condition for maximum $\pmue$}
\label{appendix_a}
 In order to find the condition of the maximum of the oscillation probability in Eq.\ \ref{eq:prob}, we first make the following observation. 
 The second term inside the square bracket of Eq.\ \ref{eq:prob} 
 is proportional to $\alpha$, and thus is suppressed by a factor of $\mathcal{O}(10^{-2})$ compared to the first term inside the square bracket. 
 Hence for simplification, $\pmue$ is maximum when the first 
 term is maximum, - which in turn implies the maximality of 
 $\sin(1-A)\Delta_{31}/(1-A)$. 
 We further note that,
 \begin{align}\label{eq:max_derivation}
&\frac{\sin(1-A)\Delta_{31}}{1-A} \nonumber \\
&\simeq (1+A)\sin \Delta_{31} - A\Delta_{31}\cos\Delta_{31} \nonumber \\
&= \sqrt{(1+A)^{2}+A^{2}\Delta_{31}^{2}}\sin(\Delta_{31}-\gamma),
\nonumber \\
&\simeq (1+A)\sin(\Delta_{31}-\gamma),
 \end{align}
 where $\gamma = \tan^{-1}\big(\frac{A\Delta_{31}}{1+A}\big)$, 
 and we have neglected terms of the order of $\mathcal{O}(A^{2})$ in the above approximation. 
 But with the same approximation,
 $\gamma \simeq A\Delta_{31}(1-A)$, and thus Eq.\ \ref{eq:max_derivation} further reduces to, 
 \begin{align}\label{eq:max_derivation1}
\frac{\sin(1-A)\Delta_{31}}{1-A} \simeq \sin[(1-A)\Delta_{31}] + A\sin[(1-A)\Delta_{31}]. 
 \end{align}
 If $A$ is not too large (\ie, not close to 1), this becomes   maximum when $\sin[(1-A)\Delta_{31}]$ is maximum. 
 This justifies Eq.\ \ref{eq:osc_max}.
\section*{Acknowledgments}
NF, MM, KS acknowledge support from the grant NRF-2022R1A2C1009686. 
MM thanks Joachim Kopp for helpful discussions regarding flux normalization in GLoBES. 
NF and MM thank Alberto Gago for helpful discussions on the article. 
NF and MM acknowledge the warm hospitality of Brookhaven National Laboratory, where this work was initiated.
GY would like to thank Praveen Kumar for the useful discussion.
This work reflects the views of the authors and not those of the THEIA or DUNE collaboration.

 \bibliographystyle{JHEP}
\bibliography{reference}
\end{document}